\documentstyle[prd,preprint,aps]{revtex}
\begin{document}
\draft
\title{Fermion Mass Hierarchy without Flavour Symmetry}
\author{Chih-Lung Chou}
\address{Stanford Linear Accelerator Center, Stanford, CA 94309}
\address{Applied Physics Department, Stanford University, Stanford,
 CA 94309}
%\date{\today}
\maketitle
\begin{abstract}
We discuss a supersymmetric grand unified model which has gauge group
 SU(5) $\times$ SU(5) $\times$ SU(5), with matter fields transforming
 asymmetrically under different gauge SU(5) groups.  We observe that the
 gauge structure of the model leads to approximate texture zero
 structures in fermion mass matrices and a natural hierarchy in the
 Yukawa couplings. The proton lifetime is estimated to be larger than
 $10^{34}$ years in this model. As in more conventional supersymmetric
 GUT models with product gauge groups, this model possesses no tensor 
fields with rank higher than 2, so that it might arise from a level 1
 string construction. 
\end{abstract}

\section{Introduction}

The Standard Model (SM) is now considered to be completely successful in
 describing the physical world up to the weak scale.  However, it
 requires some 18 parameters which are input by hand to fit the
 experiment data. Most of these undetermined parameters reflect our lack
 of understanding of flavour physics.  The SM provides no explanation of
 why there is a mass hierarchy among the fermion masses and no
 explanation of the CKM angles. It seems that Nature includes some
 classification which goes beyond the structure of the SM \cite{Peskin}.
 Thus, in order to solve these puzzles, we have to go beyond the SM. The
 Minimal Supersymmetric Standard Model (MSSM) \cite{MSSM} has considered
 as one of the possible extended theories beyond the SM. Despite of its
 success in providing true gauge coupling unification \cite{coupling}, it
 also has flavour problems \cite{flavor} at least as severe as those in
 the SM. The fermion mass hierarchy is still left unexplained in the MSSM
 framework. Even worse, new problems such as Flavour Changing Neutral
 Currents \cite{FCNC} occurs. 

Many solutions have been proposed for the flavour problem, either within
 a supersymmetric framework \cite{review,TYPICAL} or in
 non-supersymmetric theories. Most of these attempts assume that some
 flavour symmetries, gauged or global, exist above the grand unification
 scale $M_x$. The flavour symmetries typically restrict the possible
 Yukawa coupling terms in the superpotential and provide textures and
 hierarchy patterns of the fermion mass matrices
 \cite{GUTflavor,AU1,TYPICAL}.  This idea is often combined with that of
 grand unification.  For example, one can introduce higher rank tensors
 such as the 126 in the $SO(10)$ grand unified theories (GUT's)
 \cite{126} and the 45 in the $SU(5)$ GUT's \cite{45} in order to create
 specific textures in the quark and lepton mass matrices.  Theoretically,
 there is nothing wrong with introducing high rank tensors. However, the
 affine level 1 constructions in string theory does not allow
 string-derived GUT's having tensor fields with rank higher than 2
 \cite{stringGUT}.  This result makes the ordinary SUSY $SU(5)$
 \cite{SU5} and $SO(10)$ \cite{SO10} GUT theories difficult to obtain
 from the affine level 1 constructions in string theory.  In response to
 this situation, Barbieri et. al. \cite{GG} pointed out that extending
 the GUT gauge group to be $G \times G$, where G could be some GUT groups
 such as the $SU(5)$ or the $SO(10)$ group,  makes it possible to
 construct GUT models which break the product gauge groups down to the SM
 gauge group without introducing high rank tensor fields.  The GUT gauge
 group in these theories could be broken by fields which carry
 fundamental and antifundamental representations under two different
 gauge groups. For examples, the (5,$\bar 5$) and the ($\bar 5$,5) can
 break the $SU(5) \times SU(5)$ GUT theories down to the SM gauge group. 
 The same logic applies to theories with gauge group 
$G \times G \times G$. Furthermore, as pointed out by Barbieri et. al
 \cite{GGG}, if we choose
 to have each family of matter fields transforming under different gauge
 group G, then a flavour theory could be constructed without the need for
 an explicit flavour symmetry group.

In this paper, we follow the idea of using
 $SU(5) \times SU(5) \times SU(5)$ as the SUSY GUT gauge group.  However,
 instead of symmetrically assigning each family of matter fields
 (10+$\bar 5$) to its own gauge group $SU(5)$, we assign the matter
 fields transforming under these gauge $SU(5)$'s in an asymmetrical way. 
 In Section 2, we describe our model and suggest a suitable vacuum for
 those fields which break the GUT gauge group. A $Z_2 \times Z_3$
 discrete symmetry at the superheavy scale is introduced to suppress the
 dangerous operators as well as to obtain a weak-scale $\mu$ value in the
 model. This gives the full set of assumptions of our construction. In 
the remainder of the paper, we show that these assumptions lead to many
 interesting consequences.  In Section 3, we derive the fermion mass
 matrices and demonstrate a mass hierarchy which follows from the gauge
 structure of our model.  We show that our model can account for the
 observed fermion mass matrices and CKM angles.  In Section 4, we discuss
 proton decay in this model.  The proton lifetime predicted in this model
 is consistent with the limit set by the SuperKamiokande experiment.  In
 Section 5,  we present some conclusions.    

\section{The Model}

Our model is based on the SUSY GUT gauge group $SU(5)_1$ $\times$
 $SU(5)_2$ $\times$ $SU(5)_3$. We identify the SM gauge group
 $SU(3)_C \times SU(2)_L \times U(1)_Y$ as lying in a diagonal $SU(5)$
 subgroup of above product group. To break the GUT gauge group down to
 the SM $SU(3)_C$ $\times$ $SU(2)_L$ $\times$ $U(1)_Y$, we require the
 exotic Higgs fields $T_1$, $T_2$ and $T_3$ in the representations
 $(1,5, {\bar 5})$, $({\bar 5},1, 5)$ and $(5,{\bar 5},1)$. We will find
 it useful to add two more multiplets, $\Sigma$ in the $(1,5, {\bar 5})$
 and $\bar \Sigma$ in the $(1,{\bar 5}, 5)$.  We assign the three $10$'s
 of $SU(5)$ to the three different $SU(5)$ groups and we associate the 
$5$ and $\bar 5$ Higgs fields with different groups.  However, we assign
 two $\bar 5$ matter multiplets to the same $SU(5)$.  The complete set of
 assignment is shown in Table 1.  According to the assignment in Table 1,
 there is already some interesting physics at the level of lower 
dimension operators.  The ordinary $\mu$ term

\begin{equation} 
\mu H {\bar H} \label{eqn:ordmu}
\end{equation}

\noindent is forbidden from appearing in the fundamental Lagrangian by
 gauge invariance. The leading contribution to the $\mu$ term potentially
 comes from high dimension operators in the superpotential and will be
 analyzed further in the later of this section.

As one can see from the table, this model contains no fields in the 
adjoint representation, and no fields with rank higher than 2. All of
 these fields can appear in a string construction with the gauge group
 realized at the affine level $k=1$ \cite{stringGUT}. The breaking of the
 GUT gauge group can be accomplished by the vacuum expectation values 
(VEV's) of the fields $T_1$, $T_2$, $T_3$, $\Sigma$ and $\bar \Sigma$. 
 The symmetry-breaking ground state could be either by a stablized 
tree-level superpotential or by effects of a strongly coupled SUSY gauge 
theory.   Here, before discussing an explicit potential, we would like to
 propose a possible vacuum which can break the gauge group 
$SU(5)_1 \times SU(5)_2 \times SU(5)_3$ down to the SM gauge group 
$SU(3)_C$ $\times$ $SU(2)_L$ $\times$ $U(1)_Y$. We assume that the 
symmetry is broken in two steps.  First, $SU(5)_1 \times SU(5)_3$ is 
broken to the diagonal subgroups by an expection value of $T_2$.

\begin{equation}
\langle T_2 \rangle = \Lambda_2 \cdot \mbox{diag}(1,1,1,1,1)
\label{eqn:VEVT2}
\end{equation}

\noindent Then the remaining symmetry $SU(5)_{D31} \times SU(5)_2$ is 
broken to the $SU(3)_C \times SU(2)_L \times U(1)_Y$ by the expection 
values of $T_1$ and $\Sigma$.
 
\begin{eqnarray}
\langle T_1 \rangle = \Lambda_1 \cdot \mbox{diag}(0,0,0,1,1) 
\label{eqn:VEVT1} \\
\langle \Sigma \rangle = \Omega \cdot \mbox{diag}(1,1,1,0,0) 
\label{eqn:VEVOmega} 
\end{eqnarray}

\noindent Finally, the remaining fields get their expection values along 
the $SU(3)_C \times SU(2)_L \times U(1)_Y$ direction.  Only relatively 
small hierarchies between these scales are needed to produce large 
hierarchies in the quark mass matrices.  We will show this in Section 3. 
The complete pattern of VEV's consistent with the symmetry breaking 
pattern just described is:

\begin{eqnarray}
\langle \Sigma \rangle &=& \Omega \cdot \mbox{diag}(1,1,1,0,0) 
\hspace{1cm} \langle {\bar \Sigma} \rangle = {\bar \Omega}  \cdot 
\mbox{diag}(1,1,1,a,a) \nonumber \\
\langle T_1 \rangle &=& \Lambda_1 \cdot \mbox{diag}(0,0,0,1,1) 
\hspace{1cm} \langle T_3 \rangle = \Lambda_3 \cdot \mbox{diag}(1,1,1,s,s)
 \nonumber \\
\langle T_2 \rangle &=& \Lambda_2 \cdot \mbox{diag}(1,1,1,1,1) + 
\Lambda_3 \cdot \mbox{diag}(0,0,0,b,b)   
\label{eqn:VEV}
\end{eqnarray}
%%
%%%%%

\noindent Due to the $SU(5)_1$ D-term condition, the VEV 
$\langle T_2 \rangle$ will receive a correction of  order $O(\Lambda_3)$.
  The constants $a$, $b$ and $s$ are assumed to be nonzero and would be 
determined by minimizing the potential. We will show below that the zeros
 in $\Sigma$ and $T_1$ can be exact, up to the point where SUSY is 
spontaneously broken. As in conventional GUT models, we also requuire a 
discrete symmetry to forbid dangerous operators such as  $H {\bar 5_1}$,
 $H {\bar 5_3}$, $10_3{\bar 5_3}{\bar 5_1}$ and $T_1 T_3 H {\bar 5_2}$ 
in the tree-level superpotential.  Specifically, we assume a 
$Z_2^{matter} \times Z_3$ symmetry 

\begin{eqnarray}
Z_2^{matter}: (10_1,10_2,10_3,{\bar 5_1},{\bar 5_2},{\bar 5_3}) 
&\longrightarrow& -1(10_1,10_2,10_3,{\bar 5_1},{\bar 5_2},{\bar 5_3})
 \nonumber \\
Z_3: (H, {\bar H}, \Sigma , {\bar \Sigma}, 10_3) &\longrightarrow&
 (H, {\bar H}, \Sigma , {\bar \Sigma}, 10_3) \nonumber \\
(T_1, T_3, 10_2, {\bar 5_2}) &\longrightarrow& e^{i2\pi/3}
(T_1, T_3, 10_2, {\bar 5_2}) \nonumber \\
(T_2, 10_1, {\bar 5_1}, {\bar 5_3}) &\longrightarrow& e^{i4\pi/3}
(T_2, 10_1, {\bar 5_1}, {\bar 5_3}) 
\label{eqn:Z2Z3}
\end{eqnarray}

\noindent The dangerous dimension five operators that could make the 
proton decay too rapidly are also suppressed by the $Z_3$ symmetry. We 
will discuss this in Section 4.  We now discuss the spectra of Higgs 
masses and the $\mu$ paramater.  Applying the $Z_2^{matter} \times Z_3$ 
symmetry, we can easily write all possible leading terms up to 
dimension 10 level that are bilinear in $H$ and $\bar H$.
\begin{eqnarray}
W_{H {\bar H}} &=& \Sigma H{\bar H}\{ 1+\frac{\Sigma{\bar \Sigma}}{M^2}+
\frac{\Sigma T_2T_3}{M^3}+\frac{(\Sigma{\bar \Sigma})^2}{M^4} +
\frac{\Sigma^5 +\Sigma^2 T_1^3}{M^5}+\frac{(\Sigma {\bar \Sigma})
(\Sigma T_2T_3)}{M^5} \nonumber \\
&+&\frac{(\Sigma{\bar \Sigma})^3 + (T_1{\bar \Sigma})^3 + 
(\Sigma T_2 T_3)^2}{M^6}+ \sum_{k=0}^5 \frac{1}{M^{5-k}}{\bar \Sigma}^k 
(T_2T_3)^{5-k} \}\nonumber \\
&+&T_1 H {\bar H} \{ \frac{(T_1{\bar \Sigma})^2}{M^4}+
\frac{\Sigma^3 T_1^2+ T_1^5 + T_3^5 + (T_1 {\bar \Sigma})(T_1T_2T_3)}
{M^5} \nonumber \\
&+& \frac{(T_1T_2T_3)^2}{M^6} \}
\label{eqn:HHbar}
\end{eqnarray}       

\noindent From the vacuum state described in Eq.\ (\ref{eqn:VEV}), not 
all terms in Eq.\ (\ref{eqn:HHbar}) would have non-zero contributions to
 the Higgs triplet mass and the $\mu$ value. The Higgs triplets get a 
superheavy mass $\Omega$ which is shown to be of $O(10^{16})$ GeV in the 
next section. The leading terms that give $\mu$ a nonzero value are 

\begin{eqnarray}
\mu \approx \langle T_1 \lbrack  \frac{(T_1{\bar \Sigma})^2}{M^4}+
\frac{\Sigma^3 T_1^2+T_3^5+(T_1 {\bar \Sigma})(T_1T_2T_3)}{M^5}+ 
\frac{(T_1T_2T_3)^2}{M^6} \rbrack \rangle.
\label{eqn:mu}
\end{eqnarray}

\noindent Eq.\ (\ref{eqn:mu}) is highly suppressed by $1/M^4$. When we 
estimate the various paramaters in the next section, we will see that 
$\mu$ obtains a weak-scale $\mu$ value. 

It is important to ask whether the pattern of VEV's that we have 
considered in the Eq.\ (\ref{eqn:VEV}) can follow from a tree-level 
superpotential. There is an example of a superpotential that can lead to
 this structure which incorporates the constraints of 
$Z_2^{matter} \times Z_3$ symmetry. 

%%%%% Equation
%%
\begin{eqnarray}
W(\Sigma, \bar \Sigma, T_1, T_2, T_3)&=&\frac{1}{M^3}Y_1(\Sigma^3 T_1^2  
- \phi_1^{5}) + \frac{1}{M^3}Y_2(T_2^5 - \phi_2^5) + \frac{1}{M^3}Y_3 
\Sigma ^4 T_1 \nonumber \\
&+& \frac{1}{M^3}Y_4 (\Sigma^2 T_1^3) + \frac{1}{M}Y_5(\phi_1^3 - 
\Lambda^3)+ \frac{1}{M}Y_6(\phi_2^3 - \Lambda_2^3) \nonumber \\
&+& \sum_{i=1}^6 X_i Y_i^2 + Y_7 \Sigma {\bar \Sigma} + \frac{1}{M}Y_8 
\Sigma T_2 T_3 + MX_7 Y_7 + MX_8Y_8 \nonumber \\
&+&\frac{A_1}{M}(\Sigma {\bar \Sigma})^2 +\frac{B_1}{M^2}(\Sigma {\bar 
\Sigma})(\Sigma T_2 T_3) +\frac{B_2}{M^3}(\Sigma T_2 T_3)^2 \nonumber \\
&+&\sum_{i,j \ge 0}^{3} C_{ij} \frac{(T_1{\bar \Sigma})^i (T_1T_2T_3)^j 
(T_2^5)^{3-i-j}}{M^{12-3i-2j}}
\label{eqn:TTB}
\end{eqnarray}
%%
%%%%%

\noindent Here $A_i$, $B_i$ and $C_{ij}$ are understood as the 
unspecified coefficients and M is the super-high scale. The gauge 
singlets $\phi_i$, $Y_i$ and $X_i$ are needed to produce the following 
constraints

\begin{eqnarray}
\langle \Sigma^3 T_1^2 \rangle = \Lambda^5, \hspace{1cm} \langle T_2^5 
\rangle = \Lambda_2^5  \label{eqn:constr1} \\
\langle \Sigma^4 T_1 \rangle = 0, \hspace{1cm} \langle \Sigma^2 T_1^3 
\rangle = 0. \label{eqn:constr2} 
\end{eqnarray}

\noindent These lead to the zero texture patterns in the VEV's of 
$\Sigma$ and $T_1$. The F-term conditions from the superpotential 
(\ref{eqn:TTB}) as well as the D-term conditions of the GUT gauge groups 
would determine the possible vacua of this model. The $SU(5)_3$ D-term as
 well as the $SU(5)_2$ D-term conditions could force the scales 
$\Lambda_1$ and $\Omega$ to have approximately equal value 
$\Lambda_1 \approx \Omega$ if $\Lambda_1$ is much larger than 
$\Lambda_3$ and 
$\bar \Omega$. Typically, solving for the minima of a potential would 
give rise to many discretely degenerate vacua. This is generic to most 
SUSY GUT theories \cite{GGG,GG,stability} if a tree-level superpotential 
is responsible for breaking the GUT gauge group.
	
The above Higgs triplet-doublet splitting mechanism is similiar to the 
sliding-singlet mechanism \cite{SSmechanism}.  The Higgs triplets and 
doublets split when the field $\Sigma$ get superheavy VEV's in its 
$SU(3)$ block, while keeping vanishing VEV's in its $SU(2)$ block. This 
description applies to the theory before supersymmetry breaking.  It is 
a well-known difficulty of the sliding-singlet mechanism that SUSY 
breaking effects could bring corrections to the VEV of $\Sigma$ and may 
destory the gauge hierarchy \cite{gaugehierarchy}. We will now argue 
that this is not a problem in our model.

To be explicit, the problem resides in \cite{gaugehierarchy} is that the 
low energy effective singlet field $\Sigma_s$ that comes from the field 
$\Sigma$ couples to the superheavy heavy triplets in $H$ and $\bar H$.  
If we turn on SUSY breaking effects, this would give rise to one-loop 
tadpole graphs which induce the following two terms in the low energy 
effective theory.

\begin{eqnarray}
c_1 m_g^2 M_G \Sigma_s + h.c. \label{eqn:term1}\\
c_2 m_g M_G F_{\Sigma_s} + h.c. \label{eqn:term2}
\end{eqnarray}

\noindent Here $m_g$ represents the gaugino mass and $M_G$ represents 
the GUT mass scale. These terms shift the VEV's of $\Sigma$ and $T_1$. 
 Adding Eq.\ (\ref{eqn:term1}) to the effective theory, the piece of the
 potential that could shift the VEV's of $\Sigma$ in its $SU(2)$ block is
 given by

\begin{eqnarray}
V&=&({\vert \langle H \rangle \vert}^2 + {\vert \langle {\bar H} 
\rangle \vert}^2) {\vert \langle \Sigma \rangle \vert}^2 + {\vert 
\langle \frac{1}{M^3}\Sigma^4 T_1 \rangle \vert }^2 + {\vert \langle 
\frac{1}{M^3} (\Sigma^3 T_1^2 - \Lambda^{5}) \rangle \vert}^2 \nonumber
 \\
&+&{\vert \langle \frac{\Sigma^2 T_1^3}{M^3} \rangle \vert}^2 +{\vert 
\langle \Sigma {\bar \Sigma}+ M X_7 \rangle \vert}^2 + {\vert \langle 
\frac{1}{M}\Sigma T_2T_3+ M X_8 \rangle \vert }^2 + c_1 m_g^2 M_G 
\Sigma_s \nonumber \\
&+& \cdots .
\label{eqn:shift}
\end{eqnarray} 

\noindent Inserting $\Sigma \longrightarrow \Sigma + \Delta \Sigma$ into
 Eq.\ (\ref{eqn:shift}), we find the possible shift of the $SU(2)$ VEV's
 put

\begin{equation}
\Delta \Sigma_2 \stackrel{<}{\sim} \frac{c_1 m_g^2 M_G}{2 {\vert 
\frac{4}{M^3} \Sigma^3 T_1 \vert}^2} \sim O(10^2) \mbox{GeV} 
\label{eqn:delta}
\end{equation}   

\noindent For the same reason, the VEV's in the $SU(3)$ block of the 
field $T_1$ could also receive an order $10^2$ GeV correction. As one
 can see from Eq.\ (\ref{eqn:delta}), the shift of the VEV 
$\langle \Sigma \rangle$ in its $SU(2)$ block is bounded and would not
 destroy the gauge hierarchy.  The same strategy can be applied to the
 term in Eq.\ (\ref{eqn:term2}). After eliminating the auxiliary field 
$F_{\Sigma_s}$, this term gives a potential of the form

\begin{eqnarray}
\vert Y_7 {\bar \Sigma} + \frac{Y_8 T_2 T_3}{M}+ \frac{4 Y_3 \Sigma^3
 T_1}{M^3} + \frac{3 Y_1 \Sigma^2 T_1^2}{M^3}+\frac{2 Y_4 \Sigma T_1^3}
{M^3}+ H {\bar H}+ c_2 m_g M_G + \cdots \vert^2.
\label{eqn:shift2}
\end{eqnarray}

\noindent This modification can shift the VEV's of the singlets $Y_1$, 
$Y_3$ by an amount of order $10^9$ Gev or shift the VEV's of the singlets
 $Y_7$ and $Y_8$ by an amount of order of $10^4$ and $10^5$ GeV; this 
gives a small correction to the potential which is consistent with the 
hierarchy. 

In this section, by extending the GUT group from the commonly used 
$SU(5)$ group to $SU(5)_1 \times SU(5)_2 \times SU(5)_3$, we are allowed
 to solve the Higgs triplet-doublet splitting problem and give $\mu$ a
 weak-scale value. It seems that having the H and ${\bar H}$ transform 
under different $SU(5)$ gauge groups gives a natural mechanism for 
solving these problems. Thus it is well-motivated to introduce product 
groups like $SU(5) \times SU(5)$ or the $SU(5) \times SU(5) \times SU(5)$
 group as potential SUSY GUT gauge groups.

\section{The Fermion Mass Matrices}

Now we examine the structure of the Yukuwa couplings in our model. Just 
as we constructed the terms bilinear in Higgs fields, it is 
straightforward to write the terms bilinear in quark and lepton fields. 
 For the up quark masses, there are terms that apply $H10_i10_j$ to 
various combinations of the GUT-level Higgs fields.

\begin{eqnarray}
W_{up} &=& H \{10_310_3 + \frac{T_1}{M}10_2 10_2 + \frac{T_2^2}{M^2}10_3
 10_1 + \frac{T_1 T_3}{M^2} 10_1 10_1 +  \nonumber \\
&+&10_310_2(\frac{T_1^2 \Sigma}{M^3}) +10_210_1(\frac{T_1^2 T_2^2 \Sigma}
{M^5})+ \cdots \} 
\label{eqn:Wup}
\end{eqnarray}

\noindent In the above superpotential $W_{up}$, we list only the leading 
terms to various combinations bilinear in fields $10_i$. The omitted 
terms in Eq.\ (\ref{eqn:Wup}) represent possible next to leading order 
combinations.  For the down quark and lepton masses, we find terms that 
include ${\bar H}10_i{\bar 5_j}$ contracted with various combinations of 
the GUT-level Higgs fields. 

\begin{eqnarray}
W_{down-lepton} &=& {\bar H}\{\frac{T_1}{M}10_3 {\bar 5_3} 
+ (\frac{T_1(\Sigma {\bar \Sigma})}{M^3}+\frac{T_1(\Sigma T_2T_3)}{M^4} )
 10_3{\bar 5_1} \nonumber \\
&+& 10_3{\bar 5_2}[\frac{(T_1{\bar \Sigma})(\Sigma^2 T_3)}{M^5})] + 
\frac{T_3T_2+{\bar \Sigma}}{M^2}10_2{\bar 5_3} + \frac{T_3}
{M}10_2{\bar 5_2} \nonumber \\
&+& \lbrack\frac{T_3T_2}{M^2}+\frac{\bar \Sigma}{M}\rbrack10_2{\bar 5_1}
 + \frac{T_1T_2^2}{M^3}10_1{\bar 5_3} + \frac{T_2T_1}{M^2}10_1{\bar 5_2}
 + \frac{T_1T_2^2}{M^3}10_1{\bar 5_1} \nonumber \\
&+& \cdots \} 
\label{eqn:Wdl}
\end{eqnarray}

\noindent  We have defined the two matter fields $5_1$ and $5_3$, which 
have the same gauge and $Z_2^{matter} \times Z_3$ quantum numbers, so 
that the first term of Eq.\ (\ref{eqn:Wdl}) contains only $5_3$, and 
$5_1$ is the orthogonal linear combination.  We have ignored all the 
coefficients that could appear in front of each coupling term in Eq.s\ 
(\ref{eqn:Wup}) and (\ref{eqn:Wdl}).  Terms such as 
$T_2^4 10_110_1 H/{M^4}$ and 
${\bar \Sigma}^4 10_3 {\bar 5_3} {\bar H}/{M^4}$ in the superpotentials 
$W_{up}$ and $W_{down-lepton}$ are not listed because they are the 
higher-order contributions to the entries of the fermion mass matrices. 
We will see this point much clearly in the later discussion of this 
section. However, as we will see in Section 4, the term 
$({T_2^4}/{M^4})10_110_1H$ cannot be ignored in the discussion of the 
proton decay in the model. As is typical in GUT theories based on 
$SU(5)$ unification \cite{SU5}, the up-type fermion masses are seen to
 be unrelated to the down- and lepton-type fermion masses.

According to Eq.s\ (\ref{eqn:Wup}) and (\ref{eqn:Wdl}), only the top 
quark will receive a weak-scale mass. All other fermion masses arise 
from nonrenormalizable couplings and thus are suppressed by powers of 
$1/M$.  These powers, together with the various VEV's in Eq.\ 
(\ref{eqn:VEV}), lead to a hierarchy of Yukawa couplings.  To exhibit 
this hierarchy, define the small paramaters $\rho={\Omega}/{M}$, 
${\bar \rho}={\bar \Omega}/{M}$, $\xi_1={\Lambda_1}/{M}$, 
$\xi_2={\Lambda_2}/{M}$ and $\xi_3={\Lambda_3}/{M}$.  Then the leading 
contributions to each element of the Yukawa matrix is  
 
\begin{eqnarray}
(Up)_{{\bar u_i}u_j}&=&\left(
   	\begin{array}{ccc}
   	s \xi_1\xi_3 & 0  & \xi_2^2 \\
	 \xi_1^2 \xi_2^2 \rho  & \xi_1 & \xi_1^2 \rho \\
	\xi_2^2 & 0  & 1      
	\end{array}
   \right) \label{eqn:up} \\
(Down)_{{\bar d_i}d_j}&=&\left(
	\begin{array}{ccc}
	\xi_1\xi_2^2 &  \xi_2\xi_3+{\bar \rho} &\xi_1 \rho {\bar \rho} \\
	 \xi_1\xi_2  & \xi_3    & 0 \\
	 \xi_1\xi_2^2 &\xi_2\xi_3+{\bar \rho} & \xi_1  
	\end{array}
     \right) \label{eqn:down}\\
(Lepton)_{{\bar e_i}L_j}&=&\left(
	\begin{array}{ccc}
	\xi_1\xi_2^2 &(s)\xi_2\xi_3+(a){\bar \rho}  & \xi_1 \rho 
{\bar \rho} \\
	 \xi_1\xi_2  & (s)\xi_3    &0 \\
	 \xi_1\xi_2^2 &(s)\xi_2\xi_3+(a){\bar \rho} & \xi_1  
	\end{array}
     \right) \label{eqn:lepton}
\end{eqnarray}

\noindent From the above mass matrices, a approximate texture zero 
structure \cite{ansatz,SO10operator} would be presented in the up quark
 mass matrix after determining the scale ratios. The down quark mass 
matrix and the lepton mass matrix are identical, except that the $(1,2)$,
 $(2,2)$ and $(3,2)$ entries of the lepton mass matrix have different 
coefficients.  These differences are due to the VEV patterns of $\langle 
{\bar \Sigma} \rangle$ and $\langle T_3 \rangle$.  

Before making further comments on the mass matrices, we would like to
 point out that if the introduced $Z_3$ symmetry is disabled in the
 model, then the forbidden terms such as $({{\bar \Sigma}^2}/{M^2} + 
{T_2^2T_3^2}/{M^4})10_310_2H$, 
$({T_1T_3^3}/{M^4}+{T_1^3T_2^2}/{M^5})10_210_1H$ and 
$(T_1^2 T_3/M^3)10_3{\bar 5_2}{\bar H}$ will give additional 
contributions to $W_{up}$ and $W_{down-lepton}$.  We list these terms 
in the Appendix, Eq.\ (\ref{eqn:AWup}) and (\ref{eqn:AWdl}).  These new
 terms show the same hierarchy in powers of the small paramaters 
$\rho$, $\bar \rho$ and $\xi_i$.  In other words, the mass hierarchy 
is merely determined by the gauge structure but not by the global 
discrete symmetry in the model. 

Since this model cannot predict the coefficients for the coupling terms 
in superpotential, we assume these to be of order O(1) and ignore all 
coefficients in the above mass matrices.  The zero entries in the up 
quark mass matrices are only approximate and could be replaced by those
 ignored subleading terms in Eq.\ (\ref{eqn:Wup}). In fact, by the 
estimation made in later in this section, these ``zeros'' are such small
 numbers that they should be smaller than $10^{-11}$. Therefore, we can 
just ignore them in the later discussion. 

Although we do not know the coupling term coefficients, however, we can 
still extract some interesting points from Eq.\ (\ref{eqn:up} - 
\ref{eqn:lepton}). First, this model requires a low value of 
$\tan \beta$ because the top Yukawa coupling is much larger than the 
bottom Yukawa coupling.  

We also observe that because of the VEV structures of 
$\langle {\bar \Sigma} \rangle$ and $\langle T_3 \rangle$, the terms 
$({T_2T_3}/{M^2}+{\bar \Sigma}/{M}) 10_2 {\bar 5_1} {\bar H}$ and  
$({T_3}/{M})10_2{\bar 5_2}{\bar H}$ have different  contributions to 
the down-quark mass matrix and the lepton mass matrix.  It has long 
been a problem for $SU(5)$ grand unification that the mass relation 
$m_l=m_d$ at the GUT scale cannot be obeyed for all three generations.  
Georgi and Jarlskog \cite{Georgi,stringSO10}, proposed a solution which 
has been used in models of SUSY $SO(10)$ grand unification 
\cite{SO10op1}. Despite the successful experimental data fitting in 
their model, the low energy mass relation ${m_s}/{m_d}=25.15$ predicted 
in their model is two standard deviations away from the average value 
obtained by sum rule and chiral perturbation methods 
\cite{sumrule,SO10operator}.  Our scheme does not give a definite 
prediction for the mass relations, but it does give some required extra
 freedom.  For example, if the coefficient $s$ is taken to be 3, then we
 obtain the GUT scale mass relations

\begin{eqnarray}
m_{\tau} &=& m_b \label{eqn:btau} \\
m_{\mu} &\approx& 3 m_s \label{eqn:mus} 
\end{eqnarray} 

\noindent These GUT mass relations could lead to acceptable 
$m_b/m_{\tau}$ and $m_{\mu}/m_s$ mass relations 
\cite{SO10operator,ansatz} at the weak scale.

 A specific choice of the parameters that gives an acceptable 
representation of all of the experimental data on fermion masses is the 
following: 

\begin{eqnarray}
\frac{m_c}{m_t} &\sim& \xi_1 \sim O(10^{-2}),\label{eqn:ct} \\
\frac{m_u}{m_c} &\sim& s \cdot \xi_3  \sim O(10^{-2})\label{eqn:muc} \\
\frac{m_s}{m_b} &\sim& \frac {\xi_3}{\xi_1} \sim O(10^{-1}) 
\label{eqn:sb}\\
\frac{m_e}{m_{\mu}} &\sim& \frac{\xi_1 \xi_2^2}{s \xi_3}+
\frac{ a \xi_1 \xi_2 ({\bar \rho}+\xi_2\xi_3)}{(s \xi_3)^2} 
\sim O(10^{-2}) \label{eqn:emu}\\
\frac{m_d}{m_s} &\sim&  \frac{\xi_1 \xi_2^2}{\xi_3}+ \frac{\xi_1 
\xi_2 ({\bar \rho}+\xi_2 \xi_3)}{\xi_3^2} \sim O(10^{-1}) \label{eqn:ds} 
\end{eqnarray}

\noindent The above relations allow us to choose the scale ratios as 

\begin{eqnarray}
\xi_1 &\sim& \rho \sim \frac{1}{3} \times 10^{-2} \label{eqn:T1}\\
\xi_2  &\sim& 3 \times 10^{-2}\label{eqn:T2} \\
\xi_3 &\sim& \frac{1}{3} \times 10^{-3} \label{eqn:T3} \\
{\bar \rho} &\sim& \frac{1}{2}\times 10^{-4}.\label{eqn:sbar}
\end{eqnarray}

\noindent From the $\mu$ value equation in (\ref{eqn:mu}), it can be 
easily checked that these values would give rise to a weak-scale 
$\mu$ value.  Based on the given scale ratios, we can also estimate the 
CKM mixing angles $s_{12}$, $s_{23}$ and $s_{13}$ by

\begin{eqnarray}
s_{12} : s_{23} : s_{13} \sim \frac{\xi_1 \xi_2}{\xi_3} : \frac 
{{\bar \rho}+\xi_2 \xi_3}{\xi_1} : \xi_2^2 \sim O(10^{-1}) 
: O(10^{-2}) : O(10^{-3}),\label{eqn:CKM}
\end{eqnarray}

\noindent which is consistent in order of magnitude with the experimental
 data. The GUT-group breaking scales are now determined to have the 
relation $\Lambda_2 > \Lambda_1 > \Lambda_3$.  This confirms the 
breaking pattern described in Section 2.

According to the scale ratio estimations, there are approximate texture 
zero structures in the fermion mass matrices.

\begin{eqnarray}
(Up)_{{\bar u_i}u_j}&=&\left(
   	\begin{array}{ccc}
   	s \xi_1\xi_3 & 0 & \xi_2^2 \\
	 0  & \xi_1    & \xi_1^2 \rho  \\
	\xi_2^2 & 0 & 1      
	\end{array}
   \right) \label{eqn:up2} \\
(Down)_{{\bar d_i}d_j}&=&\left(
	\begin{array}{ccc}
	\xi_1\xi_2^2 & \xi_2\xi_3+{\bar \rho} & \xi_1 \rho {\bar \rho} \\
	\xi_1\xi_2   & \xi_3    & 0 \\
	\xi_1\xi_2^2 & \xi_2\xi_3+{\bar \rho} & \xi_1  
	\end{array}
     \right) \label{eqn:down2}\\
(Lepton)_{{\bar e_i}L_j}&=&\left(
	\begin{array}{ccc}
	\xi_1\xi_2^2 & (s)\xi_2\xi_3+(a){\bar \rho} & \xi_1 \rho 
{\bar \rho} \\
	\xi_1\xi_2   & (s)\xi_3    & 0 \\
	\xi_1\xi_2^2 & (s)\xi_2\xi_3+(a){\bar \rho} & \xi_1  
	\end{array}
     \right) \label{eqn:lepton2}
\end{eqnarray}

\noindent The zero entries are only approximate and represent values 
smaller than $10^{-10}$. Unlike the case in conventional SUSY flavor 
models \cite{flavor,ansatz,SO10operator,SU5}, these texture zeros are 
the natural outcome of the gauge structure as well as the scale ratios 
given in the model. In other words, they could arise without flavour 
symmetry. 

In this section, we have estimated the possible scale ratio values 
needed to obtain acceptable fermion mass structures. The GUT gauge group 
$SU(5)_1 \times SU(5)_2 \times SU(5)_3$ would undergo a two-step breaking
 down to the SM group $SU(3)_C \times SU(2)_L \times U(1)_Y$.  The SM 
gauge couplings unify at the scale of $10^{16}$ Gev if we take the 
superheavy scale $M$ to be the reduced Planck scale.  The Higgs triplets
 $H_c$ and $\bar H_c$ would obtain GUT scale masses of order of $10^{16}$
 Gev due to the superpotential term $\Sigma H {\bar H}$. Although we did 
not discuss the possible threshold effects \cite{threshold} caused by 
those exotic Higgs fields as well as the heavy Higgs triplets, it is 
quite interesting that we find naturally a hierarchical pattern for the 
fermion mass matrices.

\section{Proton Decay}
We have already introduced a $Z_2^{matter}$ symmetry to disable all 
dangerous dimension three and four operators in Section two.  However, 
since we find Higgs triplet masses of order $10^{16}$ GeV, there is a 
danger that dimension five operators which violate baryon and lepton 
number could cause fast proton decay \cite{Proton}.  A dimension five 
operator in the superpotential could lead to proton decay if it has the 
form

\begin{eqnarray}
\frac{\lambda}{M^*}Q_1Q_1Q_2L_i.\label{eqn:proton}
\end{eqnarray}

\noindent Here $Q_i$ and $L_i$ represent the $i^{th}$ generation of the 
quark and lepton multiplets respectively, $M^*$ represents some high 
scale, and $\lambda$ is the coupling constant. This operator leads to 
proton decay through the mode $p \longrightarrow K^+ {\bar \nu}$.  The 
current experiment data have already set the limit 
${\lambda}/{M^*} \stackrel{<}{\sim} 10^{-24} \mbox{GeV}^{-1}$ with the 
naturalness assumption that all squark/slepton masses are no larger than 
1 TeV \cite{Proton,murayama}.  In principle, operators of the form of 
Eq.\ (\ref{eqn:proton}) could arise from integrating out particles with 
GUT-scale masses or directly from the higher-dimension operators in the 
original Lagrangian.  In the Appendix, we analyze these higher-dimension 
operators and show that they are highly suppressed by powers of $1/M$ 
due to the gauge structure as well as the $Z_3$ symmetry of the model. 
Therefore, the main contributions to proton decay in the model will come 
from heavy Higgsino exchange processes.

Since the VEV $\langle T_1 T_3 \rangle$ has vanishing contribution to 
color triplets, the potentially leading term $({T_1T_3}/{M^2})10_110_1H$ 
cannot participate in the heavy Higgsino exchange processes. The same 
logic also applies to the terms such as $({T_1}/{M})10_210_2H$, 
$({T_1T_2}/{M^2})10_1{\bar 5_2}{\bar H}$ and 
$({T_1T_2^2}/{M^3})10_1{\bar 5_1}{\bar H}$. Therefore, by taking the 
quark mixings into account, the leading terms in the superpotential 
that contribute to the dimension five operators in 
Eq.\ (\ref{eqn:proton}) are the following:

\begin{eqnarray}
10_310_3H \label{eqn:Q33} \\
\{ \frac{T_2^2}{M^2} \}10_110_3H  \label{eqn:Q13} \\
\{ \frac{T_2^4}{M^4} \}10_110_1H  \label{eqn:Q11} \\
\{ \frac{\bar \Sigma}{M}+\frac{T_2T_3}{M^2} \}10_2 
{\bar 5_1}{\bar H}   \label{eqn:QB21} \\
\{\frac{T_3}{M} \}10_2 {\bar 5_2}{\bar H}   \label{eqn:QB22} 
\end{eqnarray}

\noindent From Eq.s\ (\ref{eqn:Q33} - \ref{eqn:QB22}), the 
leading dimension five operators that come from integrating out 
heavy Higgs triplets are shown in Fig.\ (\ref{fig:figa}) and Fig.\ 
(\ref{fig:figb}).  We find that Fig.\ (\ref{fig:figa}) should dominate 
the proton decay in the model with the decay mode 
$p \rightarrow K^{+} {\bar \nu_{\mu}}$. There are two contributions
 to Fig.\ (\ref{fig:figa}), with coupling strengths

\begin{eqnarray}
\frac{\lambda}{M^*} &\sim& \frac{1}{M_{H_c}} \times 
\frac{\langle T_2^4 \rangle}{M^4} \times \frac{\langle T_3 \rangle}{M} 
\stackrel{<}{\sim} 10^{-25} \mbox{Gev}^{-1} \label{eqn:strength} \\
\frac{\lambda}{M^*} &\sim& \frac{\sin{\theta_{13}}}{M_{H_c}} \times 
\frac{\langle T_2^2 \rangle}{M^2} \times \frac{\langle T_3 \rangle}{M} 
\stackrel{<}{\sim}10^{-25} \mbox{Gev}^{-1}. \label{eqn:strength2} 
\end{eqnarray}

\noindent In Eq.\ (\ref{eqn:strength}), the factor 
$\langle {T_2^4}/{M^4} \rangle$ comes from the next leading term 
$T_2^4 10_110_1 H /M^4$ and the factor ${\langle T_3 \rangle}/{M}$ 
comes from $({T_3}/{M})10_2{\bar 5_2}{\bar H}$ in superpotential. 
In Eq.\ (\ref{eqn:strength2}), the factor $\langle {T_2^2}/{M^2} \rangle$
 comes from the term $({T_2^2}/{M^2})10_110_3H$ and $\sin{\theta_{13}}$ 
represents the mixing angle between the first and the third generation 
up-type quarks.   The above coupling strength estimations show that the 
proton lifetime in the model should be no less than $10^{34}$ years.  
This result is about 100 times longer than the current experiment limit 
\cite{data}.  It is observed to the future experiment limit that could 
be set by SuperKamiokande. 

Although there are uncertainties in determining the coefficients of the 
Yukawa coupling terms in the superpotential, however, the branching ratio
 between the $p \rightarrow K^{+} {\bar \nu_{\mu}}$ channel and the 
$p \rightarrow K^{+} {\bar \nu_{e}}$ could be definitely given by

\begin{equation}
\frac{BR(p \rightarrow K^{+} {\bar \nu_{e}})}{BR(p \rightarrow K^{+} 
{\bar \nu_{\mu}})} = (\frac{{\bar \rho}+\xi_2\xi_3}{\xi_3})^2 \sim 
10^{-2}. \label{eqn:BR}
\end{equation}

\noindent This branching ratio prediction is generic to some SUSY 
models\cite{protondecayrate} that have the down quark mass generated 
by the seesaw mechanism.  It is not clear to us how this prediction 
could be tested.

\section{Conclusion}

In this paper we have presented a supersymmetric GUT model based on the 
gauge group $SU(5)_1 \times SU(5)_2 \times SU(5)_3$. The Higgs fields 
and the matter fields are assigned to transform under the different 
$SU(5)$ groups in asymmetrical pattern. Exotic Higgs fields $\Sigma$, 
${\bar \Sigma}$, $T_1$, $T_2$ and $T_3$ are needed to break the GUT 
gauge group down to the SM gauge group 
$SU(3)_C \times SU(2)_L \times U(1)_Y$, and also to relate matter fields
 which transform under different gauge $SU(5)$'s.  The discrete global 
symmetry $Z_2^{matter} \times Z_3$ is imposed at the reduced Planck 
scale in such a way that some dangerous terms in superpotential are 
disabled and a weak-scale $\mu$ value for light Higgs doublets can be 
obtained. However, this discrete symmetry is the only flavour symmetry 
required in our scheme. The fermion mass hierarchy is a natural outcome 
of the gauge structure presented in this model.  That is, it is the 
breaking of GUT gauge group but not the breaking of flavour symmetry 
that generates the fermion mass hierarchy in our model. In Section 3, we 
have shown that realistic fermion mass matrices can be the result of 
this mechanism. The fermion mass relations and the CKM angles are 
estimated to be consistent with measured experiment data at low energy.  
The exotic Higgs fields also play important roles in predicting realistic
 down-quark and lepton mass relations. The fields $\bar \Sigma$ and $T_3$
 allow us to obtain the Georgi-Jarlskog relation between the leptons and 
down quark masses, and also more general relations that may be required 
by experiment.      

This model does not forbid the dimension five operators that could result
 in nucleon decays.  In fact, there are allowed tree-level dimension five
 operators in the superpotential.  However, these tree-level terms are 
suppressed by powers of the superheavy scale $M$ and thus are not 
important in discussing the proton decay. The proton decay in the model 
is mainly due to Higgsino-exchange processes. The dominant mode of 
proton decay in the model is the process 
$p \rightarrow K^+ {\bar \nu_{\mu}}$, the same dominant mode as in
 minimal SUSY SU(5) model.  Due to the VEV pattern of the field $T_1$,
 the leading term $({T_1T_3}/{M^2})10_110_1H$ terms in the superpotential
 does not participate in the heavy Higgs triplet exchange process and 
thus gives zero contribution to the proton decay.  The next leading 
order contributions of proton decay come from the term 
$({T_2^4}/{M^4})10_110_1H$ and quark mixing effects, which are more 
suppressed than the leading order term $({T_1T_3}/{M^2})10_110_1H$.  
Therefore, proton decay in this model is highly sensistive to changes 
of the scale ratio ${\langle T_2 \rangle}/{M}$.  The proton lifetime is
 estimated to be larger than $10^{34}$ years, depending on the exact 
${\langle T_2 \rangle}/{M}$ value and the unknown coefficients of 
coupling terms in superpotential.

Models with product SU(5) groups were originally introduced with 
motivations from string theory. Our model shows that this structure
 may be interesting in its own right as a possible explanation of the 
fermion mass spectra.

\appendix
\section{}

If the $Z_3$ symmetry is not introduced to the model, then all possible
 operators bilinear in $H$ and $\bar H$ that are up to the dimension 10
 level are given as follows:

\renewcommand{\theequation}{\Alph{section}.\arabic{equation}}

\setcounter{equation}{0}

\begin{eqnarray}
W_{H {\bar H}}&=& \Sigma H {\bar H} \{ 1+ \frac{\Sigma{\bar \Sigma}}{M^2}
 +\frac{T_1{\bar \Sigma}}{M^2}+\frac{T_1T_2T_3}{M^3}+
\frac{\Sigma T_2 T_3}{M^3} +\frac{(\Sigma{\bar \Sigma})^2}{M^4} 
+\frac{(T_1{\bar \Sigma})^2}{M^4} \nonumber \\
&+&\frac{(\Sigma{\bar \Sigma})(T_1 {\bar \Sigma})}{M^4} 
+\frac{1}{M^5}\lbrack (\Sigma{\bar \Sigma})(\Sigma T_2T_3+T_1T_2T_3) 
+(T_1{\bar \Sigma})(\Sigma T_2T_3+T_1T_2T_3) \nonumber \\
&+& \sum_{k=0}^{5}\frac{1}{M^{5-k}}{\bar \Sigma}^k (T_2T_3)^{5-k} 
+ \sum_{k=0}^{5}\Sigma^k T_1^{5-k} + T_2^5 + T_3^5 \rbrack \nonumber \\
&+& \frac{1}{M^6}[(T_1T_2T_3)^2+(T_1T_2T_3)(\Sigma T_2T_3)
+(\Sigma T_2T_3)^2 + \sum_{k=0}^{3}(T_1{\bar \Sigma})^k 
(\Sigma {\bar \Sigma})^{3-k} ] \} \nonumber \\
&+& T_1H{\bar H} \{ 1+ \frac{\Sigma{\bar \Sigma}}{M^2} 
+\frac{T_1{\bar \Sigma}}{M^2}+\frac{T_1T_2T_3}{M^3}
+\frac{\Sigma T_2 T_3}{M^3} +\frac{(\Sigma{\bar \Sigma})^2}{M^4} 
+\frac{(T_1{\bar \Sigma})^2}{M^4} \nonumber \\
&+&\frac{(\Sigma{\bar \Sigma})(T_1 {\bar \Sigma})}{M^4}
+\frac{1}{M^5}\lbrack (\Sigma{\bar \Sigma})(\Sigma T_2T_3+T_1T_2T_3) 
+(T_1{\bar \Sigma})(\Sigma T_2T_3+T_1T_2T_3) \nonumber \\
&+& \sum_{k=0}^{5} \frac{1}{M^{5-k}}{\bar \Sigma}^k (T_2T_3)^{5-k} 
+ \sum_{k=0}^{5}\Sigma^k T_1^{5-k} + T_2^5 + T_3^5 \rbrack 
+ \frac{1}{M^6}[(T_1T_2T_3)^2 \nonumber \\
&+&(T_1T_2T_3)(\Sigma T_2T_3) +(\Sigma T_2T_3)^2 +
 \sum_{k=0}^{3}(T_1{\bar \Sigma})^k (\Sigma {\bar \Sigma})^{3-k} ]  \} 
\label{eqn:AHH}
\end{eqnarray}

\noindent The leading Yukawa coupling terms that give masses to fermions
 are also listed below:

\begin{eqnarray}
W_{up} &=& H \{10_310_3 + \frac{T_1}{M}10_2 10_2 
+ \frac{T_2^2}{M^2}10_3 10_1 + \frac{T_1 T_3}{M^2} 10_1 10_1 +
  \nonumber \\
&+&10_310_2[\frac{{\bar \Sigma}^2}{M^2}+\sum_{k=0}^{3}\frac{\Sigma^{3-k}
 T_1^k}{M^3} +\frac{(T_2T_3)^2}{M^4}] \nonumber \\
&+&10_210_1[\frac{T_1T_3^3}{M^4}+\frac{T_2^2{\bar \Sigma}^2}{M^4}
+\sum_{k=0}^{3}\frac{T_1^kT_2^2 \Sigma^{3-k}}{M^5}] \nonumber \\
&+& \cdots \} \label{eqn:AWup}
\end{eqnarray}

\begin{eqnarray}
W_{down-lepton} &=& {\bar H}\{\frac{T_1}{M}10_3 {\bar 5_3}
 +\frac{T_1^2T_3+\Sigma^2 T_3}{M^3}10_3{\bar 5_2} + \frac{T_3T_2+
{\bar \Sigma}}{M^2}10_2{\bar 5_3} + \frac{T_3}{M}10_2{\bar 5_2}  
\nonumber \\
&+&10_3{\bar 5_1} ( \frac{T_1}{M}  [ \frac{ T_1{\bar \Sigma}+ \Sigma
 {\bar \Sigma}}{M^2}+\frac{T_1T_2T_3+\Sigma T_2T_3}{M^3} \nonumber \\
&+&\frac{(T_1{\bar \Sigma})^2+(T_1{\bar \Sigma})(\Sigma{\bar \Sigma})+ 
(\Sigma{\bar \Sigma})^2} {M^4} ] \nonumber \\
&+& \frac{{\bar \Sigma}^4}{M^4} ) + (\frac{T_3T_2}{M^2}+\frac{\bar 
\Sigma}{M})10_2{\bar 5_1} + \frac{T_1T_2^2}{M^3}10_1{\bar 5_3} 
\nonumber \\
&+& \frac{T_2T_1}{M^2}10_1{\bar 5_2} + \frac{T_1T_2^2}{M^3}10_1{\bar 5_1}
 + \cdots \}. 
\label{eqn:AWdl}
\end{eqnarray}

\noindent From (\ref{eqn:AWup}) and (\ref{eqn:AWdl}), a hierarchical 
and texure of fermion masses is still present in the model even without
 introducing the $Z_3$ symmetry. This can be seen by the following 
fermion mass matrices.

\begin{eqnarray}
(\mbox{Up})_{{\bar u_i}u_j}=\left(
   	\begin{array}{ccc}
   	(s) \xi_1\xi_3 & (s)\xi_1 \xi_3^3+ \xi_2^2 {\bar \rho^2} 
+ \xi_1 \xi_2^2 \rho^2& \xi_2^2 \\
	(s^2)\xi_1 \xi_3^3 +\xi_1^2 \xi_2^2 \rho & \xi_1 
& (a){\bar \rho}^2+ \xi_1 \rho^2 + \xi_1^2 \rho \\
	\xi_2^2 &{\bar \rho}^2+ \xi_1 \rho^2 + \xi_1^2 \rho  & 1      
	\end{array}
   \right) \label{eqn:Aup} 
\end{eqnarray}
\begin{eqnarray}
(\mbox{Down})_{{\bar d_i}u_j}&=&\left(
	\begin{array}{ccc}
	\xi_1\xi_2^2 & \xi_2\xi_3+ {\bar \rho} & \xi_1^2 {\bar \rho}
+\xi_1 \rho {\bar \rho}\\
	\xi_1\xi_2   & \xi_3    &  \rho^2 \xi_3 \\
	\xi_1\xi_2^2 & \xi_2\xi_3+{\bar \rho} & \xi_1  
	\end{array}
     \right) \label{eqn:Adown} \\
(\mbox{Lepton})_{{\bar e_i}L_j}&=&\left(
	\begin{array}{ccc}
	\xi_1\xi_2^2 & (s)\xi_2\xi_3 +(a){\bar \rho} &\xi_1^2 {\bar \rho}
+\xi_1 \rho {\bar \rho} \\
	\xi_1\xi_2   & (s)\xi_3    &  \xi_1^2 \xi_3\\
	\xi_1\xi_2^2 & (s)\xi_2\xi_3 +(a){\bar \rho} & \xi_1  
	\end{array}
     \right) \label{eqn:Alepton}
\end{eqnarray}

\noindent From the above matrices, the approximate texture zero 
structures will be present as a result of the gauge structure of the 
model. The up quark mass matrix (\ref{eqn:Aup}) becomes slightly 
asymmetrical due to the VEV structures given in Eq.\ (\ref{eqn:VEV}) and
 the gauge structure of this model. The coefficients $(s)$, $(s^2)$ and 
$(a)$ in the above matrices indicate the additional factors that come 
from the constants $s$ and $a$ in the VEV $\langle T_3 \rangle$ and 
$\langle \bar \Sigma \rangle$. Alltogether, these make the down quark
 and the lepton mass matrices different from each other even though they
 arise from the same superpotential $W_{down-lepton}$.

Without imposing $Z_3$ symmetry onto this model,  if we forbid possible 
dangerous dimension three and four operators by introducing 
$Z_2^{matter}$ symmetry, there could still exist some leading tree level
 operators that would mediate proton decay.

\begin{eqnarray}
(1+\frac{\Sigma{\bar \Sigma}}{M^2}+\frac{T_1 {\bar \Sigma}}{M^2}+ 
\cdots)\frac{T_3^2}{M^3}10_110_110_2{\bar 5_2}, \label{eqn:dangerous1} \\
(1+\frac{\Sigma{\bar \Sigma}}{M^2}+\frac{T_1 {\bar \Sigma}}{M^2}+ 
\cdots)\frac{({\bar \Sigma} T_2)^3}{M^7}10_110_110_2{\bar 5_2}, \\
(1+\frac{\Sigma{\bar \Sigma}}{M^2}+\frac{T_1 {\bar \Sigma}}{M^2}+ 
\cdots)\frac{T_3{\bar \Sigma}}{M^3}10_110_110_2{\bar 5_1},  \\
(1+\frac{\Sigma{\bar \Sigma}}{M^2}+\frac{T_1 {\bar \Sigma}}{M^2}+ 
\cdots)\frac{T_2^4 {\bar \Sigma}^2}{M^7}10_110_110_2{\bar 5_1},  \\
(1+\frac{\Sigma{\bar \Sigma}}{M^2}+\frac{T_1 {\bar \Sigma}}{M^2}+ 
\cdots)\frac{\Sigma T_2}{M^3}10_110_210_2{\bar 5_2},  \\ 
(1+\frac{\Sigma{\bar \Sigma}}{M^2}+\frac{T_1 {\bar \Sigma}}{M^2}+ 
\cdots)\frac{T_2^3 T_1^2 \Sigma}{M^7}10_110_110_2{\bar 5_2} \\
(1+\frac{\Sigma{\bar \Sigma}}{M^2}+\frac{T_1 {\bar \Sigma}}{M^2}+ 
\cdots)\frac{\Sigma T_3}{M^3}10_110_110_3{\bar 5_1} \\ 
(1+\frac{\Sigma{\bar \Sigma}}{M^2}+\frac{T_1 {\bar \Sigma}}{M^2}+ 
\cdots)\frac{T_2^3}{M^4}10_110_110_3{\bar 5_2} \\
(1+\frac{\Sigma{\bar \Sigma}}{M^2}+\frac{T_1 {\bar \Sigma}}{M^2}+ 
\cdots)\frac{{\bar \Sigma}^2}{M^3}10_210_310_3{\bar 5_1} \\
(1+\frac{\Sigma{\bar \Sigma}}{M^2}+\frac{T_1 {\bar \Sigma}}{M^2}+ 
\cdots)\frac{T_3{\bar \Sigma}}{M^3}10_210_310_3{\bar 5_2} \\
(1+\frac{\Sigma{\bar \Sigma}}{M^2}+\frac{T_1 {\bar \Sigma}}{M^2}+ 
\cdots)\frac{\Sigma}{M^2} 10_210_210_3{\bar 5_1}\\
(1+\frac{\Sigma{\bar \Sigma}}{M^2}+\frac{T_1 {\bar \Sigma}}{M^2}+ 
\cdots)\frac{\Sigma T_3 T_1}{M^4} 10_210_210_3{\bar 5_2}
\label{eqn:dangerous2}
\end{eqnarray}

\noindent The above non-renormalizable operators, if they exist in our
 model, would give effective dimension five operators that violate baryon
 and lepton numbers.  By the scale ratios given in Section 3, we find
 the largest two coupling strengths in the list to come from Eq.\ (A.11)
 and (A.17) 

\begin{eqnarray}
\sin\theta_{23} \sin\theta_{c} \frac{\langle \Sigma \rangle}{M^2} \sim
 \sin\theta_c \frac{\langle \Sigma T_2 \rangle}{M^3} \sim \frac{10^{-5}}
{M} &\sim& O(10^{-23}) \mbox{Gev}^{-1} \nonumber \\
&>& O(10^{-24}) \mbox{GeV}^{-1},
\end{eqnarray}

\noindent where the superheavy scale $M$ is taken to be the reduced 
Planck scale. This result would predict a proton lifetime which is about
 $10^2$ times shorter than the current experiment limit. Fortunately, if
 the $Z_3$ symmetry is introduced, some of the tree-level terms in 
Eq.s\ (\ref{eqn:dangerous1}-\ref{eqn:dangerous2}) are forbidden.  We are
 thus left with the leading tree-level terms of Eq.\ (A.7), (A.9) and 
(A.16).  

\begin{eqnarray}
\frac{(T_1 {\bar \Sigma})}{M^2}\frac{T_3^2}{M^3}10_110_110_2{\bar 5_2},
 \hspace{1cm} \frac{\lambda}{M^*} &\sim& \frac{\langle (T_1{\bar \Sigma})
 T_3^2 \rangle}{M^5} \sim \frac{10^{-13}}{M}  \\ 
\frac{(T_1 {\bar \Sigma})}{M^2}\frac{T_3{\bar \Sigma}}{M^3} 10_110_1
10_2{\bar 5_1},  \hspace{1cm} \frac{\lambda}{M^*} &\sim& \frac{\langle 
T_3 {\bar \Sigma} (T_1 {\bar \Sigma}) \rangle }{M^4}  \sim \frac{10^{-14}
}{M}   \\
\frac{T_3{\bar \Sigma}}{M^3} 10_210_310_3{\bar 5_2},   \hspace{1cm} 
\frac{\lambda}{M^*} &\sim& \sin\theta_{13}^2 \frac{\langle T_3{\bar 
\Sigma} \rangle}{M^3} \sim \frac{10^{-13}}{M}. 
\end{eqnarray}
 
\noindent These terms are much less important than the Higgsino-exchange
 processes in Eq.\ (\ref{eqn:strength}) and (\ref{eqn:strength2}).  
Therefore, they could just be ignored in discussing the proton decay 
in this model.

%%%%% Acknowledgments
%%
\subsection*{Acknowledgments}

We thank M.E. Peskin, Y. Shirman and N. Arkani-Hamed for useful 
discussions, and especially thank M. E. Peskin for reading the 
manuscript. This work is supported by Department of Energy contract 
DE-AC03-76SF00515.

%%%%% Bibliography
%%

\begin{figure}
\caption{Dimension five operators produced by integrating out heavy 
Higgs triplets. These two operators both dominate the proton decay due
 to the quark mixing.}
\label{fig:figa}
\end{figure}

\begin{figure}
\caption{Dimension five operators produced by integrating out heavy 
Higgs triplets. These two operators both contribute to the proton decay
 due to the quark mixing.}
\label{fig:figb}
\end{figure}

\begin{table}
\caption{The Field content of the model.}
\begin{tabular}{cccc}
  		& $SU(5)_1$ 	& $SU(5)_2$ 	& $SU(5)_3$ \\ \hline
$\Sigma$	&          	& 5             & $\bar 5$ \\
$\bar \Sigma$   & 	        & $\bar 5$      & 5 \\
$T_1$ 		& 		& 5		& $\bar 5$ \\
$T_2$		& $\bar 5$	&  		& 5 \\
$T_3$		& 5		& $\bar 5$	&   \\ \hline
$10_3$		&		& 		& 10\\
$10_2$		&  		& 10		&	\\
$10_1$		& 10		&		&  \\
$\bar 5_3$	&		& 		&$\bar 5$ \\
$\bar 5_2$	&  $\bar 5$ 	&	 	&  \\
$\bar 5_1$	&		& 		& $\bar 5$  \\
H		&		& 		& 5 \\
$\bar H$	& 		& $\bar 5$	& \\ \hline
\end{tabular}
\end{table}

\end{document}